\documentclass[12pt,reprint,amsmath,amssymb]{revtex4-1}
\usepackage{graphicx}
\usepackage{dcolumn}
\usepackage{bm}
\usepackage{subfigure}
\usepackage{times}
\usepackage[bookmarks=true,colorlinks=true,linkcolor=blue,
urlcolor=blue,citecolor=blue]{hyperref}

\begin{document}

\preprint{APS/123-QED}

\title{\bf Longitudinal Excitations in Bipartite and Hexagonal Antiferromagnetic Spin Lattices}

\author{Y. Xian$^1$ and M. Merdan$^{1,2}$}
 \affiliation{%
 {$^1$School of Physics and Astronomy, The University of Manchester, Manchester M13 9PL, UK}\\
$^2$Department of Physics, College of Science, University of Babylon, Babylon, Iraq }
\date{\today}

\begin{abstract}
  Based on our recently proposed magnon-density-waves using the microscopic many-body approach, we investigate the longitudinal excitations in quantum antiferromagnets by including the second order corrections in the large-$s$ expansion. The longitudinal excitation spectra for a general spin quantum number using the antiferromagnetic Heisenberg Hamiltonian are obtained for various spin lattice models. For bipartite lattice models, we find that the numerical results for the energy gaps for the longitudinal modes at $q\to0$ and the magnetic ordering wavevector $\bf Q$ are reduced by about 40-50 \% after including the second order corrections. Thus, our estimate of the energy gaps for the quasi-one-dimensional (quasi-1D) antiferromagnetic compound KCuF$_3$ is in better agreement with the experimental result. For the quasi-1D antiferromagnets on hexagonal lattices, the full excitation spectra of both the transverse modes (i.e., magnons) and the longitudinal modes are obtained as functions of the nearest-neighbor coupling and the anisotropy constants. We find two longitudinal modes due to the non-collinear nature of the triangular antiferromagnetic order, similar to that of the phenomenological field theory approach by Affleck.  We compare our results for the longitudinal energy gaps at the magnetic wavevectors with the experimental results for several antiferromagnetic compounds with both integer and non-integer spin quantum numbers, and also find good agreement after the higher-order contributions are included in our calculations.
\end{abstract}
\maketitle

\section{introduction}

The dynamics of the two-dimensional (2D) and three-dimensional (3D) quantum antiferromagnetic systems with long-ranged order at low temperature can be considered as that of a dilute gas of weakly interacting spin-wave quasiparticles
(magnons) with its density given by the quantum correction
to the classical N\'eel order \cite{Anderson1952, PhysRev.117.117, PhysRevLett.62.2313}. These magons are transverse modes with spin $S=\pm1$. The longitudinal fluctuations with spin $S=0$ present in these systems consist of the multi-magnon continuum \cite{PhysRevB.72.014413}. The question concerning long-lived, well-defined longitudinal modes in quantum antiferromagnetic systems with long-ranged order remains open. This is our main focus in this paper.

In case of the quantum antiferromagnetic systems without long-ranged order, the triplet excitation states (two transverse and one longitudinal) of the spin-1 Heisenberg chain with non-zero gap, first predicted by Haldane \cite{PhysRevLett.50.1153}, are well known. This theoretical prediction of an energy gap separating the singlet ground state from the triplet excitation states has been confirmed experimentally in the quasi-1D antiferromagnetic compounds such as CsNiCl$_3$ and RbNiCl$_3$ of the spin-$1$ \cite{PhysRevLett.56.371}. Some subsequent experimental investigations \cite{PhysRevB.50.9174,PhysRevLett.56.371, PhysRevLett.69.3571,Steiner1987,PhysRevLett.87.017201} and theoretical calculations \cite{PhysRevB.49.13235,PhysRevB.46.10854,
PhysRevLett.75.3348,PhysRevB.48.10227,PhysRevLett.62.2313} also support Haldane's conjecture. In these experiments, the temperature is high enough so the quasi-1D systems have no long-ranged magnetic order and the dynamics can be described by the 1D models. At lower temperatures, these quasi-1D antiferromagnets behave as 3D systems with long-ranged magnetic orders. The key question is whether or not the longitudinal modes survive in present of the long-ranged order and, if the answer is yes, how we describe them in general terms. In  this regard, the observation of an energy gap for the quasi-1D compound CsNiCl${}_3$ at low temperature in 1986 generated much theoretical interest \cite{PhysRevLett.56.371}. This energy gap was initially  explained by a uniaxial single-ion anisotropy but now it is widely accepted that the gapped excited state belongs to longitudinal excitation modes, first proposed by Affleck [in the quasi-1D hexagonal antiferromagnetic compounds of the $ABX_3$-type with both spin quantum number $s=1$ CsNiCl${}_3$ and RbNiCl${}_3$ \cite{Affleck1989,PhysRevB.46.8934}. A field theory approach
focusing on the spin frustrations of the hexagonal antiferromagnetic systems
has also been proposed \cite{Plumer1992}. Clearly, such longitudinal modes are beyond the usual spin-wave theory (SWT) which only predicts the transverse spin-wave excitations (magnons). There have been several theoretical
investigations in these longitudinal modes, all using the field theory approach, such as  the sine-Gordon theory for the spin-1/2 systems in bipartite quasi-1D antiferromagnetic systems treating the inter-chain couplings as perturbation \cite{Schulz1996,Essler1997}. The theoretical estimate for the energy gap of the longitudinal mode is in good agreement with the experimental results for the compound KCuF$_3$ \cite{lake2005}. More recently, a longitudinal mode was also observed in the dimerized antiferromagnetic compound TlCuCl${}_3$ under pressure with a long-range N\'eel order \cite{Ruegg2008} and tetrahedral spin system Cu$_2$Te$_2$O$_5$Br$_2$ using Raman scattering \cite{PhysRevB.67.174405} . To our knowledge, no observation of longitudinal modes in 2D or quasi-2D antiferromagnets has been reported yet.

Our theoretical investigation of the longitudinal modes in quantum antiferromagnets is based on a microscopic theory for a generic spin-$s$ Hamiltonian system \cite{Xian2006}. We identify the longitudinal excitation
states in a quantum antiferromagnet with a N\'eel-like order
as the collective modes of the magnon-density waves, which represent
the fluctuations in the magnitude of the long-range order and are
supported by the interactions between magnons. The basic idea in our analysis is similar to Feynmann's theory on the low-lying excited states of the helium-4
superfluid \cite{Feynman1954}, thereby employing the magnon-density operator $S^z$ for the antiferromagnets in place of the particle density operator for the helium-4 superfluid. Hence, the longitudinal excitation states in antiferromagnets are constructed by the $S^z$ spin operators, contrast to the transverse spin operators $S^\pm$ of the magnon states in Anderson's SWT \cite{Anderson1952}. Our preliminary calculation for the two dimensional triangular model \cite{M.Merdan2012} have been extended to the quasi-1D Hexagonal structures of CsNiCl$_3$ and RbNiCl$_3$ , where we find that our numerical results for the energy gap values at the magnetic wavevector are in good agreement with experimental results after inclusion of the high-order contributions in the large-$s$ expansion \cite{PhysRevB.87.174434}. In this article, we extend similar high-order calculations to the bipartite antiferromagnetic systems where the long-ranged order is collinear \cite{yang2011}.

We organize this article as follows. Sec. II-V cover various bipartite systems with new results from the high-order calculations. In particular, Sec. V focuses on  the tetragonal quasi-1D compound KCuF$_3$ with spin-1/2 and we find that our estimate of the minimum energy gap for the longitudinal mode after inclusion of the high-order contributions is in good agreement with the experimental result. For completeness, in Sec.~VI we include our earlier analysis \cite{PhysRevB.87.174434} for the quasi-1D hexagonal systems where there are a number of experimental results for comparison. In Sec.~VII we conclude this article by a summary and a discussion of a possible longitudinal mode in a 2D square lattice model, relevant to the parent compound La${}_2$CuO${}_4$ of the high-$T_c$ superconductors.

\section{Longitudinal Excitation in bipartite quantum antiferromagnets}

The longitudinal excitations in a quantum antiferromagnetic system with a N\'eel-like long range order correspond to the fluctuations in the order parameter. Since the quantum correction in the order parameter is given by the magnon density \cite{Anderson1952}, we identify the longitudinal modes as the magnon-density waves (MDW), supported by the interactions between the magnons. It is clear that the states of MDW may not be well defined in the high-order dimensional systems where the magnon density is very dilute and the long-range order is near the classical value with little quantum correction. However, the magnon density may be high enough in the low dimensional systems to support the longitudinal waves. In terms of microscopic many-body language, the MDW states are constructed by applying the magnon density operator $S^z$ on the ground state in a form as $S^z|\Psi_g\rangle$, similar to Feynmann's theory of the phonon-roton excitation state of the helium superfluid, where the density operator is the usual particle density operator \cite{Feynman1954,Feynman1956}.

We consider a general spin-$s$ $XXZ$ Heisenberg model on a bipartite lattice with Hamiltonian given by
\begin{equation}\label{eq1}
H=J\sum_{\langle i,j\rangle}\left[\frac12(S^+_iS^-_{j}+S^-_iS^+_{j})+AS^z_iS^z_{j}\right],
\end{equation}
where the notation $\langle i,j\rangle$ indicates the nearest-neighbor couplings only and $A$ is the anisotropy parameter $A\,(\geq1)$. The usual isotropic Heisenberg Hamiltonian is given by $A=1$. The classical ground state of Eq.~(1) is given by the N\'eel state with two alternating sublattices, one with all spin-up and the other with all spin-down. We leave the discussion of the hexagonal systems to Sec.~VI. We perform spin rotation on the spin-up sublattice by $180^\circ$ so that all spins align in the same down direction. This is equivalent to the transformation
\begin{equation}\label{eq2}
S^\mp_i\rightarrow-S^\pm_i,\quad
S^z_i\rightarrow-S^z_i,
\end{equation}
for all sites of the spin-up sublattice. The Hamiltonian~\eqref{eq1} after this transformations is given as
\begin{equation}\label{eq3}
H =-\frac14J\sum_{l\varrho}\left[(S^+_lS^+_{l+\varrho}+S^-_lS^-_{l+\varrho})+2AS^z_lS^z_{l+\varrho}\right],
\end{equation}
where $l$ runs through all $N$ sites, $\varrho$ is the nearest neighbor index vector with coordination number $z=2,4$ and 6 for linear chain, square lattice and cubic lattice respectively. The quantum ground-state of the Hamiltonian $H$ is different from the the classical N\'eel state. This difference can be quantify by a correction in the sublattice magnetization $M$ as,
\begin{equation}\label{eq4}
    M=\frac{1}{N}\sum_l\langle S_l^z \rangle_g=s-\rho,
\end{equation}
where $\langle S^z_l\rangle_g$ indicated the ground-state expectation, $s$ is the classical value, and $\rho$ is the quantum correction with the physical meaning of the magnon density. Therefore, the operator $S^z$ corresponds to the magnon-density operator, contrast to the spin-flip operators $S^\pm$ which creates or destroys magnons. In this article, we employ Anderson's SWT for our approximation of the ground state $|\Psi_g\rangle$. Anderson's SWT can be most simply formulated by expressing $S^z$ and $S^\pm$ for all sites in terms of boson operators $a^\dagger$ and $a$, as
\begin{equation}
S^z = -s+a^\dagger a,\quad
S^+ = \sqrt{2s}fa,\quad \sqrt{2s}a^\dagger f,
\end{equation}
with $f=\sqrt{1-a^\dagger a/2s}$.
For example, the linear SWT by setting $f=1$ produces the values of $\rho=0.078$ per lattice site for the spin-1/2 isotropic Heisenberg model on a simple cubic lattice, and $\rho=0.198$ per lattice site for the same model but on a square lattice. For the same model on 1D, however, SWT fails to produce the exact result of $\rho=1/2$, which represents the maximum, saturated value of the magnon density. We consider the longitudinal modes of the quantum antiferromagnetic systems in terms of the fluctuations in these magnon densities as described below.

Following Feynman as mentioned earlier, the longitudinal excitation state is approximated by applying the magnon density fluctuation operator $X_q$ to the ground state $|\Psi_g\rangle$ as
\begin{equation}\label{eq5}
|\Psi_e\rangle =X_q|\Psi_g\rangle,
\end{equation}
where $X_q$ is given by the Fourier transformation of $S^z$ operators,
\begin{equation}\label{eq6}
X_q = \frac{1}{\sqrt{N}}\sum_l e^{i\mathbf {q\cdot r}_l} S^z_l,\quad q>0,
\end{equation}
with index $l$ running over all lattice sites. The condition $q>0$ in Eq.~(7) ensures the orthogonality to the ground state. The energy spectrum for the trial excitation state of Eq.~\eqref{eq5} can be written as
\[
E(q)
 =\frac{\langle\Psi_g|\tilde X_q [H,\,X_q]|\Psi_g\rangle}{\langle\Psi_e|\Psi_e\rangle},
\]
where $\tilde X_q$ is the Hermitian of $X_q$ and where we have used the ground state equation, $H|\Psi_g\rangle =E_g|\Psi_g\rangle$. We notice that operator $S^z_l$ in $X_q$ of Eq.~(7) is a Hermitian operator, hence $\tilde X_q=X_{-q}$. By considering the similar excitation state $X_{-q}|\Psi_g\rangle$ with the energy spectrum $E(-q)=E(q)$, it is straightforward to derive \cite{Xian2007},
\begin{equation}\label{eq7}
    E(q)=\frac{N(q)}{S(q)},
\end{equation}
where $N(q)$ is given by the ground-state expectation value of a double commutator as
\begin{equation}\label{eq8}
    N(q)=\frac{1}{2}\langle[X_{-q},[H,X_q]]\rangle_g,
\end{equation}
and the state normalization integral $S(q)$ is the structure factor of the lattice model
\begin{equation}\label{eq9}
    S(q)=\langle\Psi_e|\Psi_e\rangle=\frac{1}{N}\sum_{l,l'}e^{i\mathbf q.(\mathbf r_{l}-\mathbf r_{l'})}\langle S_l^zS_{l'}^z\rangle_g.
\end{equation}
In the following sections, we apply the SWT for the approximation of the ground state $|\Psi_g\rangle$ to evaluate these expectation values.

\section{Magnon-density waves in simple lattices}

Using the Hamiltonian \eqref{eq1} and the usual spin commutation relations, it is straightforward to derive the following result for $N(q)$
\begin{equation}\label{eq10}
N(q)={zsJ}(1+\gamma_q)\,\,\tilde g_\varrho,
\end{equation}
where $\gamma_q$ is defined as usual as
\begin{equation}\label{eq11}
\gamma_q  = \frac1z\sum_\varrho e^{i{\bf q}\cdot r_\varrho}.
\end{equation}
In Eq.~(11) $\tilde g_\varrho$ is the transverse spin correlation functions defined as
\begin{equation}\label{eq12}
\tilde g_\varrho=\langle S_l^+S_{l+\varrho}^+\rangle=\Delta_\varrho-\frac{2\rho\,\Delta_\varrho+\mu_\varrho\,\delta}{2s},
\end{equation}
where
\begin{equation}\label{eq13}
\begin{split}
&\rho=\langle a_l^\dagger a_l\rangle=\frac{1}{N}\sum_q\rho_q,\quad\mu_\varrho=\langle a_l^\dagger a_{l+\varrho}\rangle=\frac{1}{N}\sum_qe^{i\bf q\cdot\varrho}\rho_q, \\
& \Delta_\varrho=\langle a_l a_{l+\varrho}\rangle=\frac{1}{N}\sum_qe^{i\bf q\cdot\varrho}\Delta_q,\quad\delta=\langle a_l a_l\rangle=\frac{1}{N}\sum_q\Delta_q,
\end{split}
\end{equation}
and where
\begin{equation}\label{eq14}
\Delta_q=\frac{1}{2}\frac{\gamma_q/A}{\sqrt{1-\gamma^2/A^2}},\quad \rho_q=\frac{1}{2}\big(\frac{1}{\sqrt{1-\gamma^2/A^2}}
    -1\big).
\end{equation}
In evaluating the correlation function $\tilde g_\varrho$ of Eq.~\eqref{eq12}, we keep up to second order in the large-$s$ expansions. The double commutator in general behaves as near the magnetic wavevector: $N(q+{\bf Q})\propto q^2,\,\,\, q\to 0,$ similar to that of the helium superfluid \cite{Feynman1954,Feynman1956}. \\

The structure factor is thus given by
\begin{equation}\label{eq15}
    S(q)=\rho+\frac{1}{N}\sum_{q'}\rho_{q'}\rho_{q+q'}+\frac{1}{N}
    \sum_{q'}\Delta_{q'}\Delta_{q+q'}.
\end{equation}
We notice that the integral in the structure factor involving the function $\gamma_{q'}\gamma_{q-q'}$ indicate the couplings between magnons, with the summation over $q$ is given by the integration
\begin{equation}\label{eq16}
\sum_q = \frac1{(2\pi)^D}\int^\pi_{-\pi} d^Dq,
\end{equation}
where $D$ is the dimensionality of the system. The longitudinal excitation $E(q)$ of Eq.~\eqref{eq7} is hence obtained by evaluating the the double commutator and the structure factor of Eqs.~\eqref{eq10}-\eqref{eq15}. We can compare the longitudinal spectrum $E(q)$ with the transverse spin-wave spectrum given by
\begin{equation}\label{eq17}
{\cal E}(q) = szJA\sqrt{1-{\gamma^2_q}/{A^2}}.
\end{equation}

\subsection{Results for linear chain (1D) Model}

The SWT breaks down for the isotropic 1D case as most integrals involving the ground-state expectation diverge. For example, $\rho\to\infty$ as $A\to1$ in the linear SWT. Furthermore, we notice that the spin-wave spectrum of Eq.~\eqref{eq17} is doublet while the exact result by Bethe ansatz \cite{Bethe1931} is triplet for the spin-1/2 model. Nevertheless, the value of Eq.~\eqref{eq17}, $J\sin q$, is not far off the exact triplet spectrum derived as \cite{desCloizeaux1962}
\begin{equation}\label{eq18}
{\cal E}^{\rm exact}(q) = \frac{\pi }2J\sin q.
\end{equation}
\begin{figure}[ht]
\centering
\includegraphics[scale=0.5]{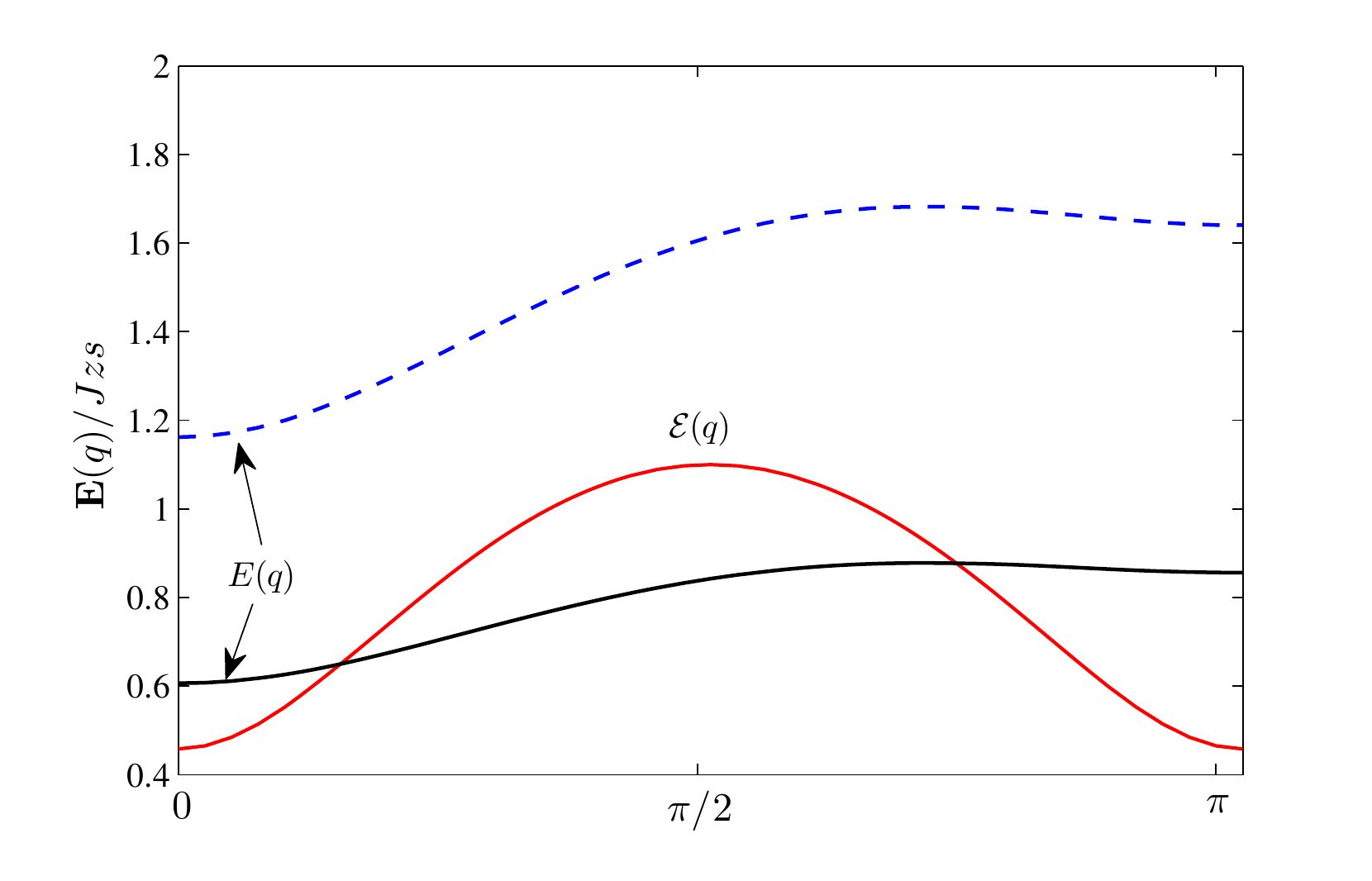}
\caption{The longitudinal excitation spectrum of  Eq.~\eqref{eq7} for the linear chain (1D), together with the spin-wave spectrum of Eqs.~\eqref{eq17} both with anisotropy $A=1.1$. The longitudinal spectra calculated from the first-order and second-order approximations are indicated by the dash and solid lines respectively.}
\label{fig1}
\end{figure}

For the longitudinal mode, we have examined the behaviors of in $N(q)$ and $S(q)$ in the isotropic limit ($A\to1$) in our earlier paper \cite{yang2011} and found that both have the same divergence as $\tilde g_1\propto -\frac1{2\pi}\ln(A-1),$  and $S(q)\to -\frac1{2\pi}{\ln(A-1)}/{\sqrt{A-1}},\quad {\rm as}\,\,\,q\to0$. Therefore, the longitudinal excitation sectrum in the isotropic limit, $E(q)\to J\sin q$, coincides precisely with spin-wave spectrum of Eq.~\eqref{eq17}. Thus, together they form the triplet excitation states, in agreement with the exact result of Bethe ansatz \cite{desCloizeaux1962}. For $A>1$,  the longitudinal excitation spectrum $E(q)$ is higher than those of the doublet spin-wave spectrum as shown in Fig.~\ref{fig1}, where we plot $E(q)$ for $A=1.1$. The gaps for the longitudinal mode $E(q)$ are about $1.16szJ$ and $1.64szJ$ at $q\to0$ and  $q=\pi$ respectively in the first order approximation of Eq.~\eqref{eq12} \cite{yang2011}. After including the second-order terms, the gap values are $0.61szJ$ and $0.86szJ$  at $q\to0$ and $q=\pi$ respectively, comparing with $0.46szJ$ of the spin-wave spectrum at the both points. We notice that the longitudinal spectrum is reduced by about half after including the second order correction, and is even lower than that of the spin-wave spectrum around $q=\pi/2$.

\subsection{Results for the square (2D) and cubic (3D) lattices}

\begin{figure}[h]
\centering
\includegraphics[scale=0.5]{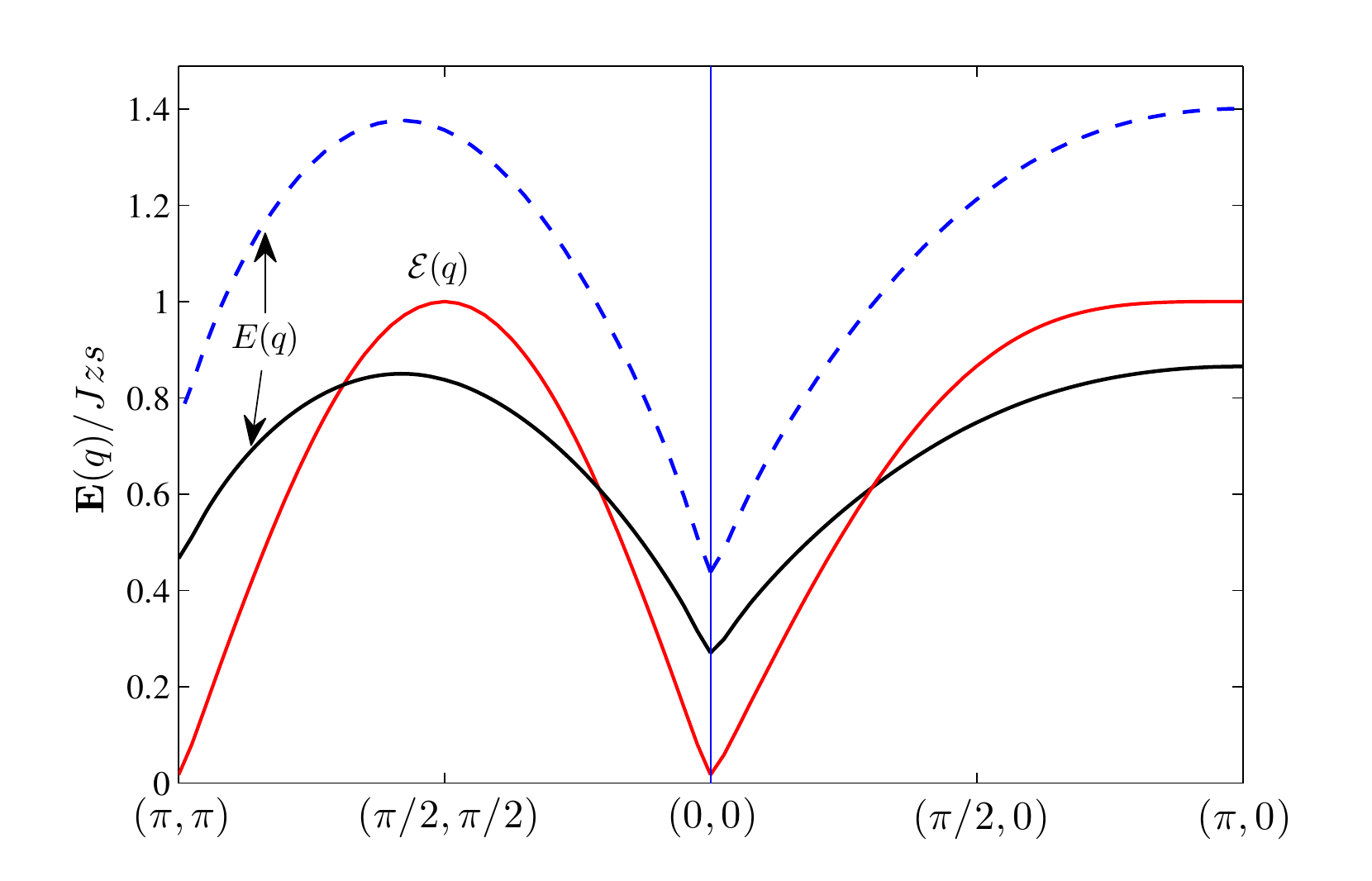}
\caption{(Color online) The longitudinal excitation spectrum of  Eq.~\eqref{eq7} for square lattice (2D) , together with the spin-wave spectrum of Eqs.~\eqref{eq17} both with anisotropy $A=1.00015$. The longitudinal spectra calculated from the first-order and second-order approximations are indicated by the dash and solid lines respectively.}
\label{fig2}
\end{figure}

For the square lattice model at the isotropic point $A=1$, we obtain, for all $\varrho$, $\tilde g_\varrho\approx0.28$ and $0.17$ for the first-order and the second-order approximations respectively. In both cases, the longitudinal excitation spectrum $E(q)$ becomes gapless due to the divergence of the structure factor in logarithmic manner, i.e. $S(q)\to-\ln q$ at $q\to0$ and the magnetic ordering wavevector ${\bf Q}=(\pi,\pi)$ as discussed in our previous paper \cite{,Xian2006,yang2011}. We also confirmed  this logarithmic behaviour of $E(q)$ in the triangular lattice model at $q\to0$ and the triangular magnetic ordering wave vector $\pm[{\bf Q}=(4\pi/3,0)]$ \cite{M.Merdan2012}. We have referred this behavior in the spectrum as "quasi-gapped" since any tiny anisotropy or finite-size effect will produce a significant gap. For example, we consider a tiny anisotropy with a value $A-1=1.5\times 10^{-4}$, which in fact is a typical value for the high-$T_c$ compound La${}_2$CuO${}_4$  \cite{PhysRevB.46.14034}, the gap values at $q\to0$ and the magnetic ordering wavevector $\bf Q$ increase to $E(q)\sim 0.44szJ$ and $E({\bf Q}) \sim 0.76szJ$ in the first order approximation, and $E(q)\sim 0.27szJ$ and $E({\bf Q}) \sim 0.47szJ$ in the second-order approximation, as shown in Fig.~\ref{fig2}.
We notice that these gap values at both points are much larger than the corresponding spin-wave gap of $0.02szJ$. At the two particular momenta $(\pi/2,\pi/2)$ and $(\pi,0)$, where $\gamma_q=0$ and the spin-wave
spectrum of Eq.~\eqref{eq17} gives the same value of $szJ$, the longitudinal spectrum $E(q)$ has slightly different values, $1.36szJ$ and $1.40szJ$ in the first-order approximation, and $0.84szJ$ and $0.87szJ$ in the second-order approximation.

For the simple cubic lattice model (3D), the numerical results for $\tilde g_\varrho$ of Eq.~\eqref{eq12} are about $0.13$ and $0.11$ in the first-order and second-order approximations respectively. The gap values at $q\to0$ and the magnetic ordering wavevector $\bf Q$ are $E(q)\sim 0.99szJ$ and $E({\bf Q}) \sim 1.40szJ$ in the first-order approximation, and $E(q)\sim 0.84szJ$ and $E({\bf Q}) \sim 1.19 szJ$ in the second-order approximation.

\section{Tetragonal quasi-1D and quasi-2D antiferromagnets}

\begin{figure}[h]
\centering
\includegraphics[scale=0.5]{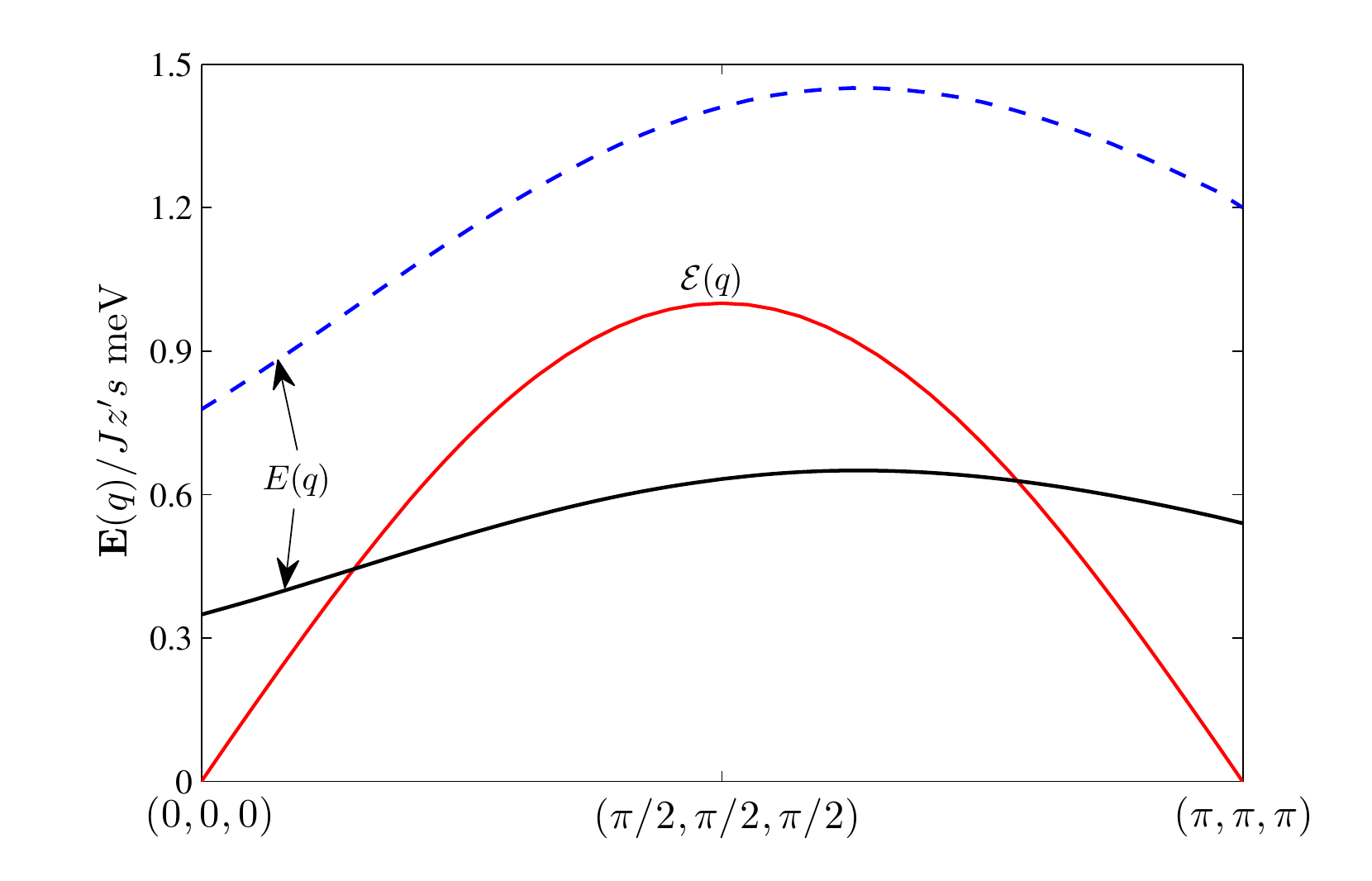}
\caption{(Color online) The longitudinal excitation spectra of Eq.~\eqref{eq7} for quasi-1D antiferromagnet with $\xi=0.05$, together with the spin-wave spectrum of Eqs.~\eqref{eq17}, both at $A=1$. The longitudinal spectra calculated from the first-order and second-order approximations are indicated by the dash and solid lines respectively.}
\label{fig3}
\end{figure}

The Hamiltonian for tetragonal quasi-1D and quasi-2D antiferromagnets is given by
\begin{equation}\label{eq19}
H=\frac12J \big(\sum^{\text{chain}}_{l,\varrho}{\bf S}_l\cdot {\bf S}_{l+\varrho}
+\xi\sum^{\text{plane}}_{l,\varrho'}{\bf S}_l\cdot {\bf S}_{l+\varrho'}\big),
\end{equation}
where $l$ runs over all the lattice sites, and $\varrho$ and $\varrho'$ are the nearest-neighbor vectors  along the chain and on the basal plane of the tetragonal structure respectively, and $\xi=J_\perp/J$ is the ratio between the  coupling constants on the basal plane $J_\perp$ and along the chain $J$. Both these coupling constants are positive for antiferromagnetic systems. The quasi-1D and quasi-2D models correspond to the cases of $\xi\ll1$ and $\xi\gg1$ respectively, whereas the simple cubic 3D model is given by $\xi=1$. The Hamiltonian operator of Eq.~\eqref{eq19} can be expressed in terms of the rotated coordinates of Eq.~\eqref{eq2}. The spin-wave spectrums and all the previous formula for the longitudinal mode with with anisotropy $A$ remain the same after the following replacements:
\begin{equation}
\begin{split}
z \to z'&=2(1+2\xi)A,\\
\quad \gamma_q\to \gamma'_q &=\frac2{z'}\left[\cos q_z+\xi(\cos q_x +\cos q_y)\right],
 \end{split}
\end{equation}
where $z'=2(1+\xi)$. In Fig.~\ref{fig3} we present the results for quasi-1D model at $A=1$ together with the spin-wave spectrum of Eq.~\eqref{eq17} for $\xi=0.05$ as an example. The gaps for $E(q)$ at $q\to0$ and $\bf Q=(\pi,\pi,\pi)$ are $0.78\,Jsz'$ and $1.20\,Jsz'$ in the first-order approximation and $0.35Jsz'$ and $0.54\,Jsz'$ in the second-order approximation.

\begin{figure}[h!]
\centering
\includegraphics[scale=0.5]{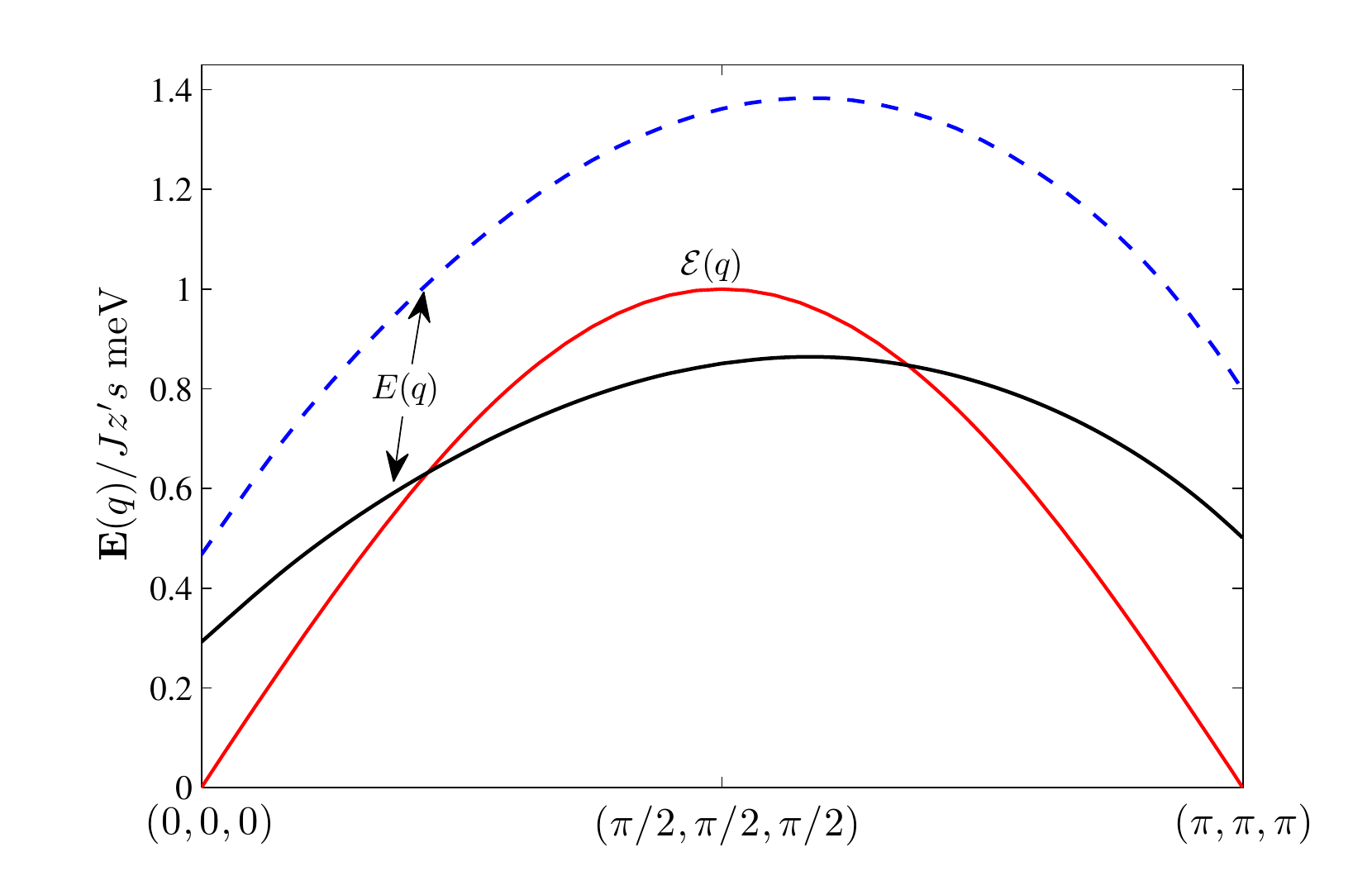}
\caption{(Color online) The longitudinal excitation spectra of Eq.~\eqref{eq7} for quasi-2D antiferromagnet with $\xi=10^3$, together with the spin-wave spectrum of Eqs.~\eqref{eq17}, both at $A=1$. The longitudinal spectra calculated from the first-order and second-order approximations are indicated by the dash and solid lines respectively.}
\label{fig4}
\end{figure}

We also presented the result for the quasi-2D system with $A=1$ and $\xi=10^3$ in Fig.~\ref{fig4}. The gaps for $E(q)$ at $q\to0$ and $\bf Q=(\pi,\pi,\pi)$ are\, $0.47\,Jsz'$ and $0.80 Jsz'$ respectively in the first-order approximation and $0.29Jsz'$ and $0.50\,Jsz'$  respectively in the second-order approximation. The longitudinal excitation spectrum at two particular momenta  $(\pi/2,\pi/2,0)$ and $(\pi,0,0)$, has slightly different values of $1.36Jsz'$ and $1.40Jsz'$ respectively in the first-order approximation, and $0.85Jsz'$ and $0.88Jsz'$ respectively in the second-order approximation. The spin-wave spectra of  Eq.~\eqref{eq17} are the same at these momentum points with the value of $Jsz'$.

\section{Quasi-1D antiferromagnetic KCuF$_3$ structure}

The KCuF$_3$ compound  of the  $s=1/2$ chains  is crystallized in the tetragonal ($Pnma$) structure with a lattice parameter $a=4.126$ ${\buildrel _{\circ} \over {\mathrm{A}}}$ and $c=3.914$ ${\buildrel _{\circ} \over {\mathrm{A}}}$ at temperature of 10K. The magnetic moment of this compound is carried by the Cu$^{2+}$ ion. Some experimental studies on this material for the magnetic excitation spectrum using inelastic neutron scattering confirm the spinon picture \cite{PhysRevB.44.12361,lake2005,PhysRevLett.70.4003,PhysRevB.52.13368}.
The magnetic interaction is represented by the  antiferromagnetic strong couplings along the chain with coupling constant $J$ and ferromagnetic weak couplings on the basal plane with couplings constant $J_\perp$. The Heisenberg Hamiltonian in terms of the lowering and raising operators with the  rotated coordinates of the chains is given by
\begin{align}\label{eq21}
H=-\frac14J &\Big[\sum^{\text{chain}}_{l,\varrho}\big(S^+_lS^+_{l+\varrho}+S^-_lS^-_{l+\varrho}+2S^z_lS^z_{l+\varrho}\big) \nonumber\\
&-\xi\sum^{\text{plane}}_{l,\varrho'}\big(S^+_lS^-_{l+\varrho'}+S^-_lS^+_{l+\varrho'}+2S^z_lS^z_{l+\varrho'}\big)\Big],
\end{align}
again where $\xi=J_\perp/J$. The spin-wave spectrum ${\mathcal E}_q$ in the linear SWT can then be obtained as
\begin{equation}\label{eq24}
{\mathcal E}_q=2Js\sqrt{\Gamma^2_q-\cos^2 q_z},
\end{equation}
where
\begin{equation}\label{eq25}
  \Gamma_q=1+2\xi(1-\gamma_q^{\rm 2D}),
\end{equation}
and
\begin{equation}
\gamma_q^{\rm 2D}=\frac12(\cos q_x+\cos q_y).
\end{equation}
The spin wave spectrum for the ferromagnetic square lattice model is recovered by setting $J=0$ and that of the linear antiferromagnetic chain is recovered by setting  $J_\perp=0$. For the longitudinal excitation, the double commutator is derived as
\begin{equation}\label{eq26}
  N(q)=-2sJ(1+\cos q_z)\tilde g_\varrho+4sJ_\perp(1-\gamma_q^{2D})\tilde g_{\varrho'}',
\end{equation}
where the correlation functions $\tilde g_\varrho$ is as defined before in Eqs.~\eqref{eq12} and \eqref{eq13} with $\Delta_q$ and $\rho_q$ given by
\begin{equation}\label{eq27}
\Delta_q =-\frac12 \frac{\cos q_z}{\sqrt{\Gamma^2_q - \cos^2q_z}}\,\,\,;\,\,\,\rho_q =\frac12 \big(\frac{\Gamma}{\sqrt{\Gamma^2_q - \cos^2q_z}}-1\big).
\end{equation}
The new correlation function $\tilde g'_\varrho$ is defined as
\begin{equation}\label{eq28}
\tilde g'_\varrho=\langle S_l^+S_{l+\varrho}^-\rangle=\mu_{\varrho}-\frac{2\rho \,\mu_{\varrho}+\Delta_{\varrho}\delta}{2s},
\end{equation}
with the parameters calculated by Eqs.~\eqref{eq13} and \eqref{eq27}.

\begin{figure}[h!]
\centering
\includegraphics[scale=0.5]{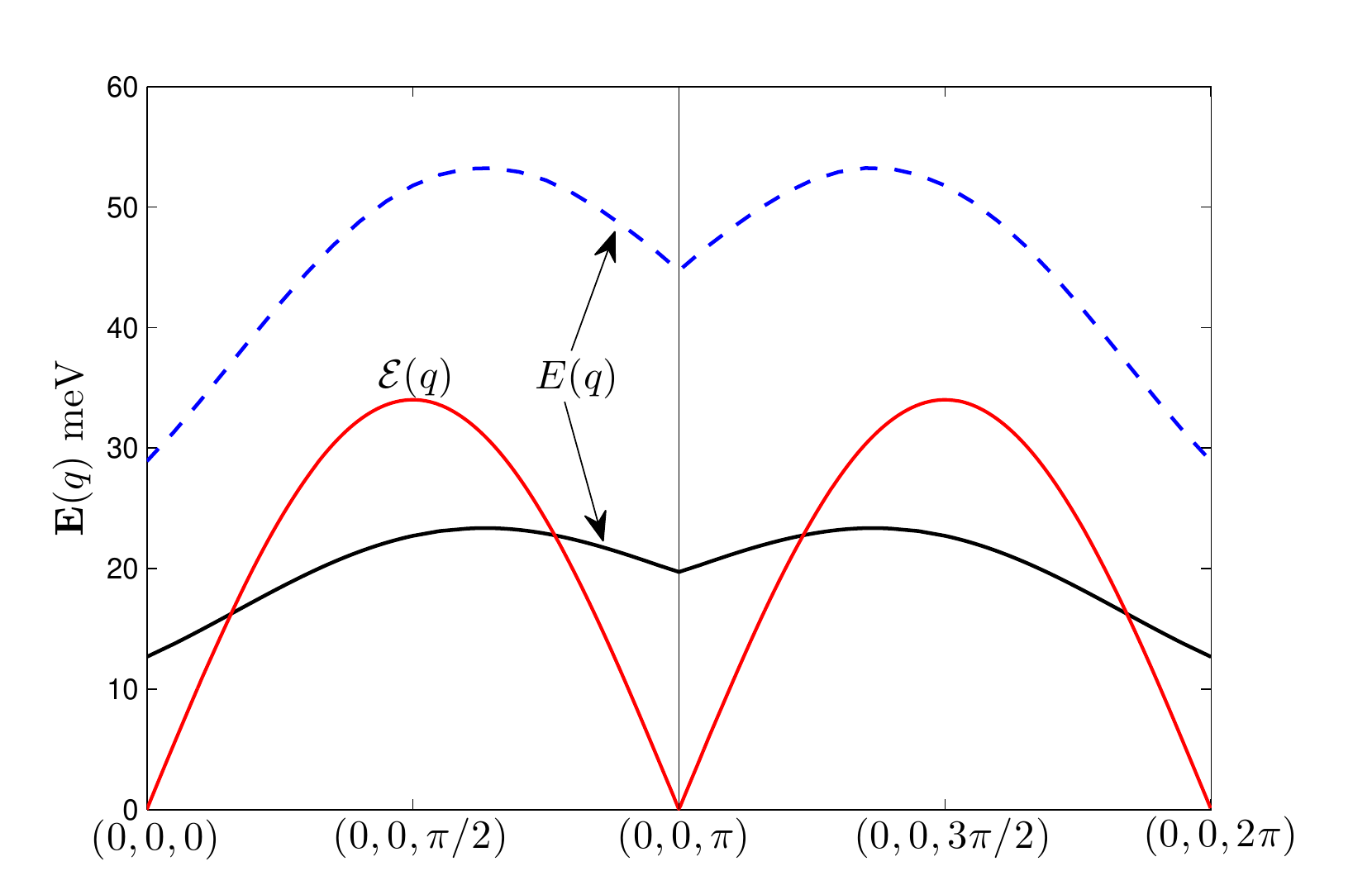}
\caption{(Color online) The longitudinal spectra of Eq.~\eqref{eq7} for KCuF$_3$ antiferromagnetic components  with $\xi\approx0.047$, together with the spin-wave spectrum of Eqs.~\eqref{eq24}. The longitudinal spectra calculated from the first-order and second-order approximations are indicated by the dash and solid lines respectively.}
\label{fig5}
\end{figure}

The structure factor can be calculated from Eq.~\eqref{eq15} using Eq.~\eqref{eq27} for $\Delta_q$ and $\rho_q$. The longitudinal excitation spectrum can thus be calculated by Eq.~\eqref{eq7}. We present the numerical results for the longitudinal spectrum in Fig.~\ref{fig5} for both the first-order and the second-order approximations, using the experimental value of the couplings constants $J\approx34$ meV and $J_\perp\approx1.6$ meV \cite{lake2005}. We notice the difference in the first-order approximation between our current results and those reported earlier in \cite{yang2011} where a wrong assumption was made in one of the integrals. Now the overall spectrum in the first-order approximation is higher than those reported earlier. In particular, the minimum energy gap at $q\to0$ for the first-order approximation is 28.9 meV and is reduced to 12.7 meV after including the second-order correction of Eq.~\eqref{eq12} and ~\eqref{eq28}. This second value of $12.7$ meV is close to the experimental value of about $15$ meV. The field theory by Essler {\it et al} produces a gap value of $17.4$ meV  \cite{Essler1997}. At the antiferromagnetic wavevector, the gaps our first-order and second-order approximations are $44.7$ and $19.7$ meV respectively.

\section{Hexagonal quasi-1D ABX$_3$-type antiferromagnetic systems}

The quasi-1D materials such as CsNiCl$_3$ crystallize in the hexagonal $ABX_3$ structure with space group $P6_3/mmc$, where $A$ is an alkaline-metal cation, $B$ is a cation of the 3$d$ group, and $X$ is a halogen anion. The magnetic ions $B$ constructs the hexagonal lattice in the $ab$ plane with adjacent spins forming angles of $\theta=2\pi/3$, and antiparallel adjacent spins along the chain, thus forming three-sublattice structure, in contrast to the bipartite systems discussed earlier. The lattice constants of CsNiCl$_3$, for example, are $a=7.14$ ${\buildrel _{\circ} \over {\mathrm{A}}}$ and $c=5.90$ ${\buildrel _{\circ} \over {\mathrm{A}}}$, and the magnetic moments are carried by Ni$^{2+}$. The superexchange interaction between $B$ (Ni$^{2+}$) ions is modeled by an $N$-spin Heisenberg Hamiltonian with a strong intrachain interaction $J$ and weak interchain interaction $J'$ such as
\begin{equation}\label{eq1.1}
    H=2J\sum_{\langle i,j\rangle}^\text{chain}\mathbf{S}_i\cdot \mathbf{S}_j+2J'\sum_{\langle i,j\rangle}^\text{plane}\mathbf{S}_i\cdot \mathbf{S}_j+D\sum_i(S_i^z)^2,
\end{equation}
where we have  added an Ising-like single-ion anisotropy term with constant $D (<0)$. Most of the intrachain couplings in $ABX_3$ compounds are antiferromagnetic such as in CsNiCl$_3$ or RbNiCl$_3$ with easy single-site anisotropy, or CsMnBr$_3$ and RbMnBr$_3$ with hard anisotropy \cite{PhysRevB.56.5373,PhysRevB.54.6327}. These intrachain couplings can also be ferromagnetic (i.e., $J<0$) as in CsNiF$_3$ \cite{PhysRevB.44.11773,PhysRevB.54.12932} or CsCuCl$_3$ \cite{Rastelli1994}. We consider only the antiferromagnetic couplings here. Therefore, the classical ground state of each linear chain along the $c$ axis (also denoted as $y$-axis)  is a N\'eel state with alternating spin-up and spin-down  alignments.

\begin{figure}
\centering
\includegraphics[scale=0.5]{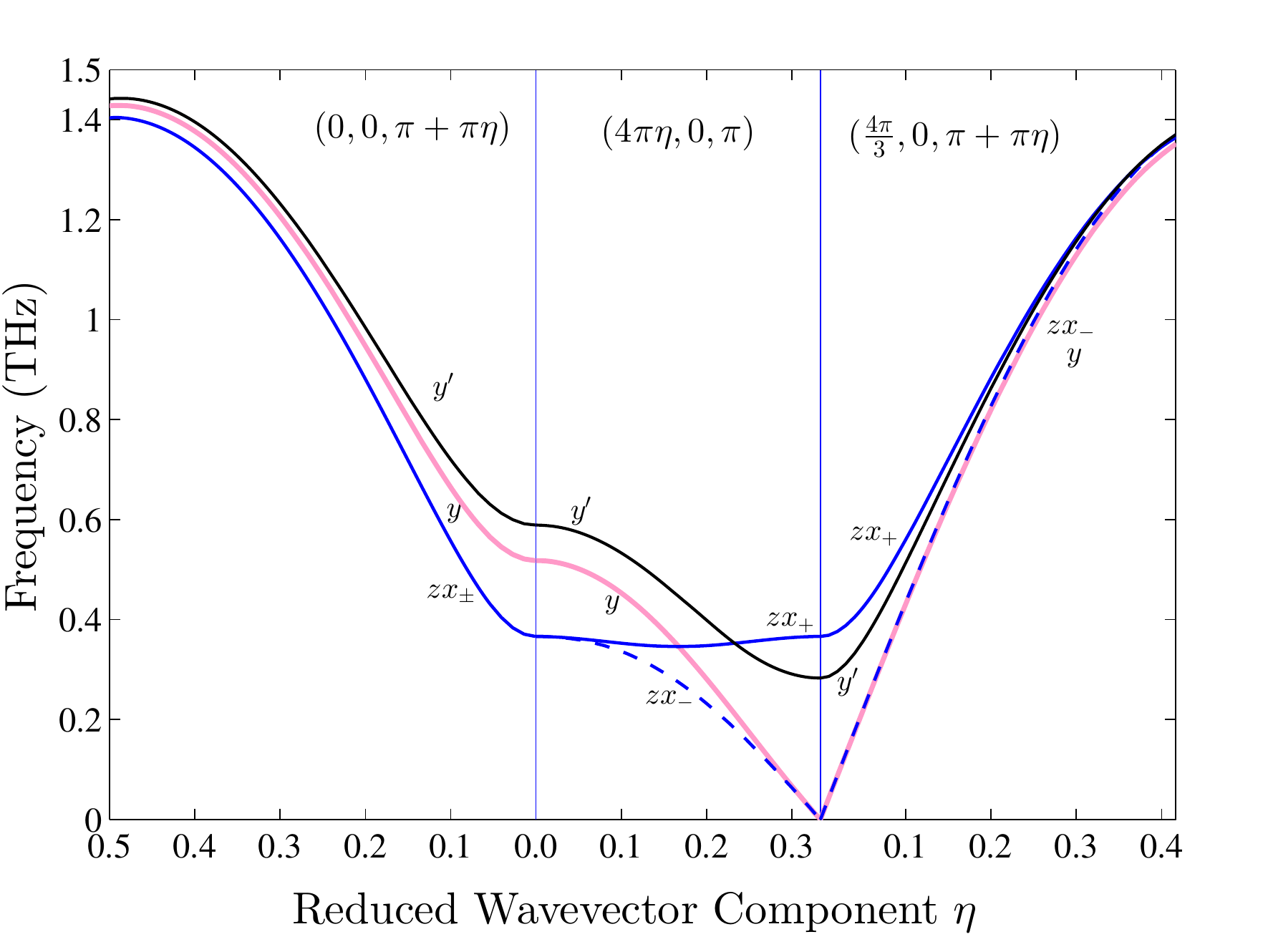}
\caption{The three spin-wave excitation spectra (in colors) for CsNiCl$_3$ with $J=0.345$, $J'=0.0054$ and $D=0$ THz, along the symmetry direction $(0,0,\pi+\pi \eta)$, $(4\pi \eta,0,\pi)$ and $(\frac{4\pi}{3},0,\pi+\pi \eta)$. Also included is the gapped $y$-mode (black, denoted as $y'$) with $D=-0.0285$ using the anisotropy term of Eq.~\eqref{eq7.1}. The solid and dash with the blue color on the lines indicate the $zx_+$-mode and $zx_-$-mode respectively.}
\label{fig3.1}
\end{figure}

As before, we employ the the spin-wave approximation for the ground state in our analysis for the longitudinal modes. We therefore first consider a spin-wave theory for the Hamiltonian~\eqref{eq1.1} based on the one-boson approach by performing two spin rotations. Firstly, we rotate the local axes of all up-spins  by $180^\circ$ so that all spins along each chain align in the same down direction. This spin rotation  is applied by using the transformation of Eq.~\eqref{eq2} to the first terms in Eq.~\eqref{eq1.1}, leaving the last two terms unchanged. The second rotation is on the hexagonal lattice of the $ab$ plane (or $xz$-plane) on the second and third terms of Eq.~\eqref{eq1.1}. Following Singh and Huse \cite{Singh1} and Miyake \cite{Miyake1992}, for every triangle of the hexagonal lattices, we rotate the local axes of two spins along the classical direction in the $xz$-plane to align with that of the third spin \cite{Chubukov1994, Chernyshev2009a}. This is equivalent to the rotation of the $i$-sites of Eq.~\eqref{eq1.1} by the following transformation
\begin{equation}\label{eq3.1}
\begin{split}
S_i^x&\rightarrow S_i^x\cos(\theta_i)+S_i^z\sin(\theta_i),\\
S_i^y&\rightarrow S_i^y,\\
S_i^z&\rightarrow S_i^z\cos(\theta_i)-S_i^x\sin(\theta_i),
\end{split}
\end{equation}
where $\theta_i\equiv\mathbf{Q_z}\cdot\mathbf{r}_i$ and $\mathbf{Q_z}=(4\pi/3,0,q_z)$ with $\mathbf Q_z$ at $q_z=\pi$ defined as the magnetic-ordering wavevector of the quasi-1D hexagonal systems. The Hamiltonian~\eqref{eq1.1} after these two transformations is given as
\onecolumngrid
\begin{equation}\label{eq5.1}
\begin{split}
H=-\frac{1}{2}J\sum\limits_{l,\varrho}^{\text{chain}}
\big[S_l^+S_{l+\varrho}^+&+S_l^-S_{l+\varrho}^-+2S_l^z
S_{l+\varrho}^z]-\frac{1}{2}J'\sum\limits_{l,\varrho'}^
{\text{plane}}\big[S_l^z
S_{l+\varrho'}^z
+\frac{3}{4}(S_l^+S_{l+\varrho'}^++S_l^-S_{l+\varrho'}^-)\\
&-\frac{1}{4}(S_l^+S_{l+\varrho'}^-+S_l^-S_{l+\varrho'}^+)
-2\sin(\theta_l-\theta_{l+\varrho'})
(S_l^zS_{l+\varrho'}^x-S_l^xS_{l+\varrho'}^z)\big]+\tilde{\cal H}^D,
\end{split}
\end{equation}
\twocolumngrid
\noindent where $l$ runs through all sites, $\varrho$ and $\varrho'$ are the nearest neighbor index vectors with coordination numbers $z=2$ along the chain and $z'=6$ on the hexagonal basal planes respectively, and $\tilde{\cal H}^D$ is the rotated anisotropy term. In order to perform the second rotation of Eq.~\eqref{eq3.1} involving rotations of the axes of the two spins to align with the axis of the third spin on the triangular planes, we rewrite the anisotropy term of the Hamiltonian~\eqref{eq1.1} in the following equivalent, suitable form
\begin{equation}\label{eq6}
 \sum_i(S^z_i)^2=\frac{1}{z'}\sum_{l,\varrho'}\left[\frac{1}{3}(S_l^z)^2+\frac
 {2}{3}(S_{l+\varrho'}^z)^2\right].
\end{equation}
The transformation of Eq.~\eqref{eq3.1} to the second term in Eq.~~\eqref{eq5.1} gives
\begin{equation}\label{eq7.1}
\begin{split}
 \tilde{\cal H}^D=&\frac{1}{z'}\sum_{l,\varrho'}[\frac{1}{3}
 D(S_l^z)^2+\frac
 {2}{3}D[(S_{l+\varrho'}^z)^2\cos^2\theta_{l+\varrho'}\\
 &+(S_{l+\varrho'}^x)^2\sin^2\theta_{l+\varrho'}
 -\cos\theta_{l+\varrho'}\sin\theta_{l+\varrho'}
 (S_{l+\varrho'}^zS_{l+\varrho'}^x\\
 &\quad\quad\quad\quad
 \quad\quad\quad\quad
 \quad+S_{l+\varrho'}^xS_{l+\varrho'}^z)].
 \end{split}
\end{equation}
After application of the usual boson transformation for the spin operators as given by Eqs.~(5), the Hamiltonian of Eq.~(31) can be expressed in terms of polynomials of the boson operators. After Fourier transformations of the boson operators with the Fourier component operators $a_q$ and $a^\dagger_q$, we have, to the order of $(2s)$ in the large-$s$ expansion
\begin{equation}
H\approx H_0+H_2,
\end{equation}
where $H_0$ is the classical energy
\begin{equation}\label{eq9.1}
    H_0=-2JNs^2-3J'Ns^2+\frac{1}{3}DNs^2
    (1+2\cos^2\theta +\frac{1}{s}\sin^2\theta)
\end{equation}
with $\theta=2\pi/3$ and $H_2$ is given by the quadratic terms in the boson operators as
\begin{equation}\label{eq10.1}
H_2=s\sum_q\big[A_qa_q^\dagger a_{-q}-\frac{1}{2}B_q(a_q^\dagger a_{-q}^\dagger+a_{q}a_{-q})\big]
\end{equation}
with constants $A_q$ and $B_q$ defined by
\begin{equation}\label{eq11.1}
\begin{split}
&A_q=4J+6J'(1+\frac{1}{2}\gamma_q)-\frac{2}{3}D(1+2\cos^2
\theta-\sin^2\theta),\\
&B_q=4J\cos q_z+9J'\gamma_q-\frac{2}{3}D\sin^2\theta,
\end{split}
\end{equation}
and $\gamma_q$ defined  by
\begin{equation}\label{eq12.1}
\gamma_q=\frac{1}{z'}\sum_{\varrho'} e^{i\mathbf {q\cdot r}_{\varrho'}}=\frac{1}{3}\Big(\cos q_x+2\cos
\frac{q_x}{2}\cos\frac{\sqrt{3}}{2}q_y\Big).
\end{equation}
The quadratic Hamiltonian $H_2$ of Eq.~\eqref{eq10.1} is diagonalized by the usual Bogoliubov transformation and can be written in terms of the new boson operators $\alpha_q$ and $\alpha^\dagger_q$ as
\begin{equation}\label{eq13.1}
H_2=\Delta H_0+\sum_q{\cal E}_q\left(\alpha_q^\dagger \alpha_q
 +\frac{1}{2}\right),
\end{equation}
where $\Delta H_0$ is the quantum correction to the classical ground state energy of Eq.~\eqref{eq9.1}
\begin{equation}
\Delta H_0=-2JNs-3J'Ns+\frac{1}{3}
DNs(1+2\cos^2\theta-\sin^2\theta),
\end{equation}
and ${\cal E}_q$ is the spin-wave excitation spectra
\begin{equation}\label{eq14.1}
{\cal E}_q=s\sqrt{A_q^2-B_q^2}.
\end{equation}
The spin-wave energy spectra with different polarizations are obtained by folding of the wavevectors. In Fig.~\ref{fig3.1}, several branches along the symmetry direction of $(0,0,\eta+1),(\eta,\eta,1)$, and $(1/3,1/3,1+\eta)$ are shown, where $\eta$ is the reduced wave vector component in the reciprocal lattice unit (r.l.u) with $q_z=(2\pi l/c)\cdot(c/2)=\pi l$, and $\gamma=1/3[\cos2\pi h+\cos2\pi k+\cos2\pi(h+k)]$. Using Eq.~(\ref{eq12.1}) the moving in the paramagnetic Brillouin zone can be written as for $q_x=4\pi\eta$ and $q_z=\pi+\pi\eta$, and the corresponding symmetry directions to those in reciprocal lattice unit are $(0,0,\pi+\pi\eta),(4\pi\eta,0,\pi)$ and $(4\pi/3,0,\pi+\pi\eta)$ respectively. The three transverse spin-wave branches are obtained from Eq.~\eqref{eq14.1} as follows. The $y$-mode has the polarization along the $y$-axis of the hexagonal lattice where the quantum fluctuation is at $q$; the other two modes are found in the $xz$-plane by translating the wavevector by a magnetic wavevector as $q\to(q\pm Q)$ and are denoted as $zx_\pm$ respectively.

As can be seen from Fig.~\ref{fig3.1}, at the magnetic wavevector $\bf Q$, the $y$-mode is gapless for zero anisotropy ($D=0$). However, as mentioned earlier, an energy gap about $0.41(2J)$ has been observed by the neutron scattering experiments for CsNiCl${}_3$ \cite{PhysRevLett.56.371}. This energy gap can be reproduced in the $y$-mode excitation by introducing an anisotropy with $D=-0.0285$ using our approximation of Eq.~\eqref{eq7.1}, also plotted in Fig.~\ref{fig3.1}. If we use the simple form of Ref.~\cite{Feile1984435} corresponding to setting $\theta=0$ in Eq.~\eqref{eq7.1}, the required anisotropy is reduced by a little more than half with the value $D=-0.0141$. Both of these values are now considered too large for CsNiCl${}_3$ which has negligible anisotropy. The conclusion is that the observed gaps are not of the transverse spin-wave spectra, but belong to the longitudinal modes, as first proposed by Affleck \cite{Affleck1989,PhysRevB.46.8934}.

Using the Hamiltonian of Eqs.~\eqref{eq5.1}, it is straightforward to derive the following double commutator with zero anisotropy (i.e. $D=0$) as
\begin{equation}\label{eq25.1}
\begin{split}
   N(q)=2sJ\sum_\varrho(1+\cos q_z)\tilde g_\varrho
   &+\frac{1}{2}J's\sum_{\varrho'}\Big[
   3(1+\gamma_q)\tilde g_{\varrho'}\\
   &-(1-\gamma_q)\tilde g_{\varrho'}'\Big],
\end{split}
\end{equation}
where $\gamma_q$ is as defined in Eq.~\eqref{eq12.1} and the transverse correlation functions $\tilde g_\varrho$ and $\tilde g'_\varrho$ are defined in Eqs.~\eqref{eq12} and \eqref{eq28} respectively, all independent of index $l$ due to the lattice translational symmetry. Also, the contribution from the three-boson operators with $\sin(\theta_l-\theta_{l+\varrho})$ (the so-called cubic term) is zero. We notice that this cubic term has been included in perturbation theory for the correction in spin-wave spectrum \cite{Miyake1992,Chubukov1994}. In the evaluation of the correlation functions $\tilde g_\varrho$ and $\tilde g_\varrho'$, we use the definition of Eqs.~\eqref{eq12}, \eqref{eq13} and \eqref{eq28} with the following expression
\begin{equation}\label{eq27.1}
\Delta_q=\frac{1}{2}\frac{B_q}{\sqrt{A_q^2-B_q^2}},\quad \rho_q=\frac{1}{2}\big(\frac{A_q}{\sqrt{A_q^2-B_q^2}}
    -1\big),
\end{equation}
with $A_q$ and $B_q$ as given before by Eqs.~\eqref{eq11.1}. The structure factor within the linear spin-wave approximation is independent of $s$, and is given by Eq.~\eqref{eq15} with the results of Eq.~\eqref{eq27.1}.

We first discuss the general behaviors of the longitudinal spectrum of Eq.~\eqref{eq7} as a function of the ratio of the two nearest-neighbor coupling constants, $\xi=J'/J$. In the limit $\xi\to0$, the Hamiltonian of ~\eqref{eq1.1} becomes the pure 1D systems; the longitudinal spectrum is gapless and identical to the doublet spin-wave spectra thus forming a triplet excitation state as discussed in Sec.~III(A). This demonstrates the limitation by the spin-wave ground-state employed, particularly when applied to the integer-spin Heisenberg chain where the Haldane gap is expected as discussed in Sec.~I. In the other limit, $\xi\to\infty$, the Hamiltonian is a pure triangular antiferromagnet with the quasi-gapped longitudinal modes as discussed in details in our previous paper \cite{M.Merdan2012} where we keep only the first order term in Eqs.~\eqref{eq12} and \eqref{eq28} in the large $s$-expansion, similar to the case of the square lattice model.

\begin{figure}
\centering
\includegraphics[scale=0.5]{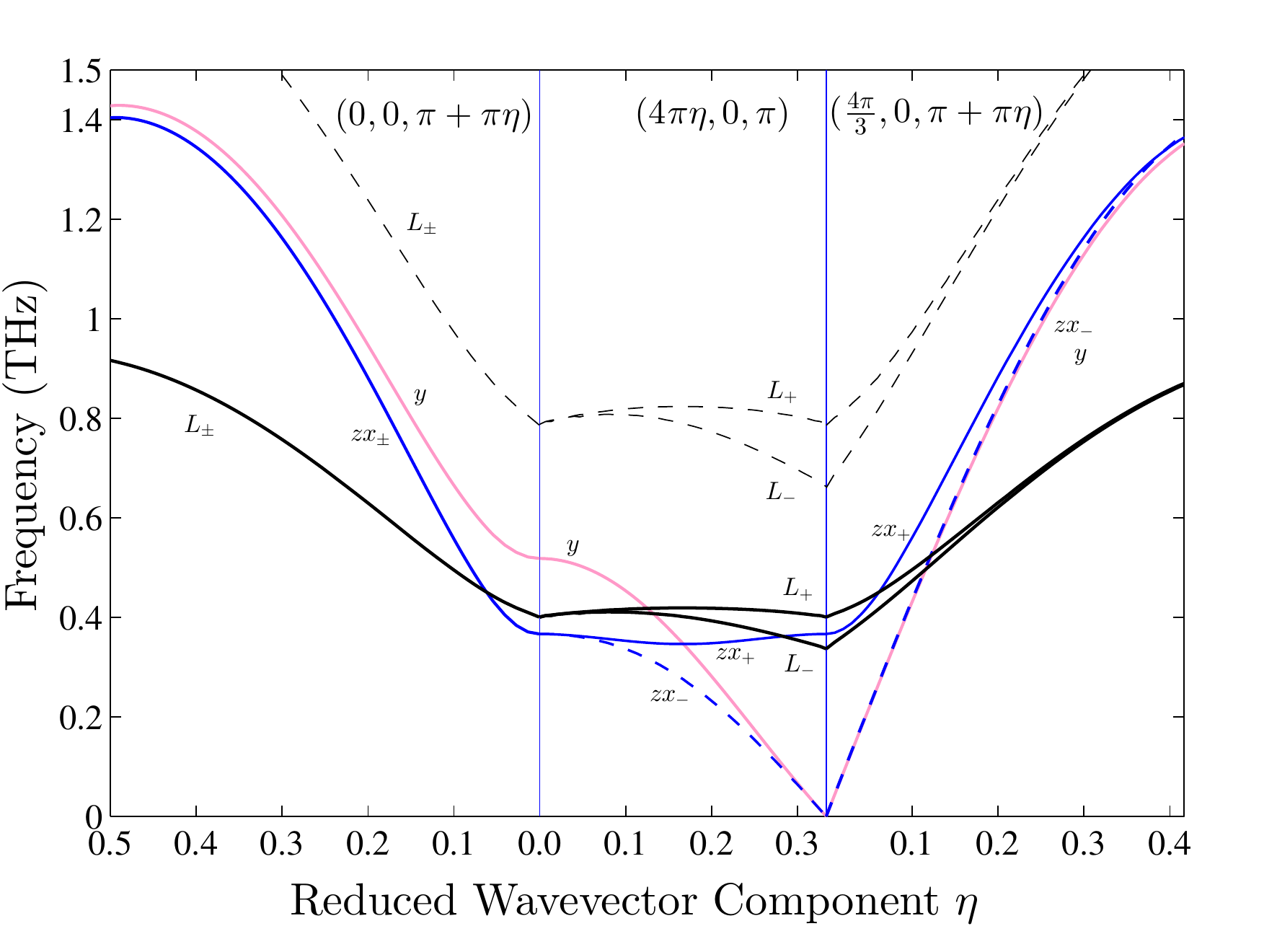}
\caption{The longitudinal modes $L_\pm$ as derived from Eq.~\eqref{eq7} together with the spin-wave $y$- and $zx_\pm$ modes as derived from Eq.~\eqref{eq14} for CsNiCl$_3$ along the symmetry direction $(0,0,\pi+\pi \eta)$, $(4\pi \eta,0,\pi)$ and $(\frac{4\pi}{3},0,\pi+\pi \eta)$. The longitudinal modes $L_\pm$ calculated from the first-order and second-order approximations are indicated by the dash and solid lines respectively.}
\label{fig4.1}
\end{figure}

For the quasi-1D materials with intermediate values of $\xi$, we expect that the spin-wave ground state is a reasonable approximation. We obtain nonzero energy gaps for the longitudinal excitation spectra of Eq.~\eqref{eq14.1}. As discussed before, following Affleck \cite{Affleck1989,PhysRevB.46.8934}, two longitudinal modes for the quasi-1D hexagonal antiferromagnets can be obtained by folding of the wavevector. We denote one as $L_-$ with the spectrum $E(q-Q)$ and the other as $L_+$ with the spectrum $E(q+Q)$. We plot these two longitudinal spectra in the first and second order approximations together with the three spin-wave spectra of Eq.~\eqref{eq14.1} in Fig.~\ref{fig4.1} near the magnetic wavector $Q$ for the compound CsNiCl$_3$. Our numerical result for the energy gap of the lower longitudinal mode $L_-$ at $Q$ is $0.96(2J)$ in the first order approximation in Eqs.~\eqref{eq12} and \eqref{eq28}. After including the second order terms the energy gap value is now $(0.49)2J$, in agreement with the experimental results of $0.41(2J)$. We also notice that the upper mode $L_+$ is higher than the $L_-$ mode by about $(0.092)2J$ at $Q$.

For the compound RbNiCl$_3$ also with $s=1$, using the exchange parameters $J=0.485$ and $J'=0.0143$ THz with a larger ratio $\xi=J'/J=0.0295$ \cite{PhysRevB.43.13331}, we obtain similar longitudinal modes as those of CsNiCl$_3$. The numerical result for the energy gap of the $L_-$  mode is 1.16 THz in the first order approximation and 0.69 THz after including the second order contributions at the magnetic wavevector. This later result is in better agreement with the experimental result of about 0.51 THz. We like to point out that there is some difficulty in fitting of Affleck's model with the experimental results for RbNiCl$_3$ \cite{PhysRevB.46.8934,PhysRevB.43.13331}.

Finally we turn to the longitudinal modes for the non-integer-spin quasi-1D hexagonal systems. The superexchange interactions in the hexagonal compound CsMnI${}_3$ can be described by the Hamiltonian of ~\eqref{eq1.1} with spin quantum number $s=5/2$ and the nearest-neighbor coupling constants $J=0.198$ and $J'=0.001$ THz and negligible anisotropy \cite{PhysRevB.43.679}. This system is very close to the pure 1D system with a very small ratio $\xi=J'/J\approx0.005$. The linear spin-wave theory may be a poor approximation for such a system. Nevertheless, with a similar analysis as before based on the spin-wave ground state, we obtain the $L_-$ mode energy gap value of $0.64$ THz  at the magnetic wavevector $Q$ in the first order approximation, and of $0.47$ THz after including the second order contributions. This later value is still much larger than the experimental value of about $0.1$ THz by Harrison \emph{et al} \cite{PhysRevB.43.679}, which was used to fit a modified spin-wave theory by Plumer and Cail\'e \cite{Plumer1992}. Clearly, for such systems as CsMnI${}_3$, we need a better ground state than that of the spin-wave theory in our analysis.

\section{Conclusion}

In this paper we have extended our high-order calculations for the longitudinal modes in the hexagonal quantum antiferromagnetic systems \cite{PhysRevB.87.174434} to a number of bipartite systems, including the the quasi-1D compound KCuF${}_3$ where good agreement in the minimum energy gap is found between the experimental result and our estimate after inclusion of the high-order contributions.

We notice that all the longitudinal modes in antiferromagnetic systems with long-ranged order have so far been observed only on a few quasi-1D systems near the critical points. This is not surprising since in these systems the magnon density is high and the longitudinal modes is well-defined and long-lived. We also notice that there is no report of longitudinal modes in 2D or quasi-2D quantum antiferromangetc systems. In particular, as we have discussed in Sec.~III(B), the longitudinal energy spectrum after inclusion of the high-order contributions, as shown in Fig.~2, is comparable to the spin-wave spectrum for the 2D square model with a tiny anisotropy same in value to the parent compound La${}_2$CuO${}_4$ of the high-$T_c$ superconductor. It will therefore be interesting to examine possible longitudinal modes in this compound near the transition to the superconducting phase when doping.

\end{document}